\documentclass[aps, prb, twocolumn, superscriptaddress]{revtex4-2}

\usepackage[utf8]{inputenc}
\usepackage{comment}

\usepackage{physics}

\usepackage[colorlinks=true, linkcolor=green, citecolor=green]{hyperref}

\usepackage{xfrac}
\usepackage{xcolor}
\usepackage{breqn}
\usepackage{amsfonts}
\usepackage{mathtools}
\DeclarePairedDelimiter\floor{\lfloor}{\rfloor}

\newcounter{phantomsub}[figure]          

\newcommand{\labelphantom}[1]{%
    \refstepcounter{phantomsub}
    \label{#1}
}

\begin{document}

\title{Effective classical potential for quantum statistical averages}

\author{Vijay Ganesh Sadhasivam}
\affiliation{ Yusuf Hamied Department of Chemistry\string, University of Cambridge\string, Lensfield Road\string, Cambridge\string, CB2 1EW\string, UK }
\author{Stuart C.~Althorpe}
\affiliation{ Yusuf Hamied Department of Chemistry\string, University of Cambridge\string, Lensfield Road\string, Cambridge\string, CB2 1EW\string, UK }
\author{Venkat Kapil}
\email{v.kapil@ucl.ac.uk}
\affiliation{ Yusuf Hamied Department of Chemistry\string, University of Cambridge\string, Lensfield Road\string, Cambridge\string, CB2 1EW\string, UK }
\affiliation{Department of Physics and Astronomy\string, University College London\string, London\string, UK.}
\affiliation{Thomas Young Centre and London Centre for Nanotechnology\string, London\string, UK\string, London\string, UK.}

\begin{abstract}
We present an effective potential that allows quantum thermal expectation values of a position-dependent observable to be estimated as a classical ensemble average of the corresponding function.
We follow the approach of Feynman and Hibbs, but perform the mean-field treatment of quantum fluctuations about the path starting point rather than the path centroid. 
Furthermore, rather than performing a full variational optimization of the potential, we explore approximate functional forms that yield a numerical robustness.
The resulting closed-form potential is exact in the classical and harmonic limits; benchmarks against exact position distributions for one-dimensional quartic, Morse, and double-well potentials, show good agreement for potentials with harmonic support.
\end{abstract}

\maketitle

\section{Introduction}
Calculating thermodynamic properties of atomistic systems from quantum mechanics is one of the goals of modern computational chemical physics.
For a quantum system of distinguishable particles with a Hamiltonian $\hat{H}$, thermodynamic properties in the canonical ensemble are determined by the partition function
\begin{equation}
Z_{\text{QM}} = \mathrm{Tr}\!\left[e^{-\beta \hat{H}}\right],
\label{eq:trace}
\end{equation}
where $\beta$ is the inverse temperature and the trace is taken over the Hilbert space on which the Hamiltonian acts.
When Eq.~\ref{eq:trace} is evaluated in a numerically exact manner by computing the eigenvalues of the Hamiltonian matrix evaluated in a chosen basis, the associated computational cost scales exponentially with the number of degrees of freedom in the system.
Consequently, such exact calculations are typically limited to systems with only a few degrees of freedom. \\

The thermodynamic properties of large systems of distinguishable particles can be calculated in a numerically exact manner using a more computationally efficient approach that follows from the imaginary-time path-integral formalism~\cite{Feynman:100771} of quantum statistical mechanics.
In this formalism, the trace in Eq.~\ref{eq:trace} can be represented as a path integral with a Euclidean action \(S\) defined on the imaginary-time interval \([0,\beta\hbar]\). 
The resulting expression for a single particle in one dimension reads
\begin{equation}
  Z_{\text{QM}} = \int \text{d}x'~ \oint_{x'} \mathcal{D} [x(\cdot)] ~e^{-{\hbar^{-1} S[x(\cdot);~\beta\hbar]}},
   \label{eq:pi}
\end{equation}
where $x(\cdot)$ is the imaginary time path, i.e., a mapping from imaginary time in $[0,\beta\hbar]$ to position space, subject to the boundary conditions $x(0)=x(\beta\hbar)=x'$, and $\mathcal{D}$ denotes the functional integration measure.
For a many-particle system living in three dimensions, the position coordinate is replaced by a $3N$-dimensional vector. 
Discretizing the path into $P$ path coordinates, spaced uniformly in imaginary time, replaces the functional integral in Eq.~\ref{eq:pi} by an ordinary $P$-dimensional integral over these coordinates~\cite{Barker1979}.
In this representation, the resulting expression is equivalent to the partition function of a classical Hamiltonian comprising $P$ replicas of the system.
The exact quantum partition function is recovered in the $P\to\infty$ limit. \\

This ensemble has proven attractive for atomistic modelling because it yields quantum thermodynamic observables only $P$ times the computational cost of a classical simulation at a given temperature~\cite{chandler_exploiting_1981, parrinello_study_1984}.
A caveat exists that $P$ increases rapidly as the temperature is lowered (as demonstrated analytically in the harmonic limit by ~\citet{markland_refined_2008}).
This has motivated the development of computational techniques that reduce the prefactor associated with $P$, including high-order splittings of the Boltzmann operator~\cite{suzuki_hybrid_1995, jang_applications_2001, perez_improving_2011, kapil_high_2016, kapil_modeling_2019}, truncated cumulant expansions of the Boltzmann operator trace~\cite{takahashi_monte_1984, poltavsky_modeling_2016, poltavsky_accurate_2020}, ring-polymer contraction~\cite{markland_efficient_2008, markland_refined_2008}, multiple time stepping~\cite{kapil_accurate_2016, marsalek_quantum_2017}, and generalized Langevin-equation thermostats that incorporate quantum nuclear fluctuations by breaking the fluctuation-dissipation theorem~\cite{ceriotti_nuclear_2009, dammak_quantum_2009, ceriotti_accelerating_2011}.
While these techniques have promoted widespread application of path-integral quantum mechanics in atomistic modeling~\cite{markland_nuclear_2018}, 
they remain asymptotically limited by the rapid growth of $P$ as the temperature is lowered. \\

The asymptotic increase in computational cost with reducing temperature arises from the increasing discretization required in imaginary time to maintain a constant error~\cite{chandler_exploiting_1981}.
It is therefore instructive to consider approaches that do not discretize the imaginary time path: for instance approximations to the Boltzmann trace in Eq.~\ref{eq:trace} that yield a quantum partition function with a functional form similar to that of a classical partition function.
This direction is motivated by the fact that classical thermodynamic properties can be estimated from a simpler ordinary integral. 
The expression for a single particle in one dimension is
\begin{align}
    Z_{\textrm{CL}} = \sqrt{\frac{m}{2\pi \beta\hbar^2}}\int  \textrm{d}x'~ e^{-\beta V(x')},
    \label{eq:classical_Z}
\end{align}
which generalizes straightforwardly to a three-dimensional many-particle system. 
There has been early interest in approximating quantum statistical mechanics using semi-classical methods or by introducing an effective classical potential that incorporates quantum effects.
Examples range from the well-known WKB approximation~\cite{wentzel_verallgemeinerung_1926, kramers_wellenmechanik_1926, brillouin_remarques_1926} to perturbative expansions of the quantum partition function in powers of $\hbar^{2}$, truncated at finite order as elucidated by \citet{wigner_quantum_1932} and \citet{kirkwood_quantum_1933} for higher dimensions.
A limitation of these approaches is that the WKB approximation is inherently one-dimensional, while perturbative methods become prohibitively expensive when incorporating higher powers of $\hbar^2$ term by term.
In this respect, the path-integral framework provides an elegant route to approximately incorporating contributions from all orders in $\hbar^2$ in the quantum partition function, by approximating the Euclidean action in Eq.~\ref{eq:pi} to a form for which the functional integral can be analytically evaluated.
This procedure yields a family of so-called quantum effective potentials,  prominently the Feynman-Hibbs~\cite{Feynman:100771} and the Feynman-Kleinert~\cite{FeynmanKleinert} potentials amongst other derivatives~\cite{giachetti1985variational,cuccoli1995effective,guillot1998quantum}, defined such that a classical partition function constructed from them approximates the quantum partition function while retaining contributions to all orders in $\hbar^{2}$. \\

Beyond the computational cost of evaluating these potentials, the most widely used quantum effective potentials exhibit a noteworthy choice: they are formulated such that they reproduce the probability distribution of the path centroid rather than the exact position distribution.
This choice poses no difficulty for partition functions in low-dimensional systems and are useful for approximating short time quantum dynamics of linear position-dependent operators~\cite{cao_new_1993, trenins_mean-field_2018,trenins_path-integral_2019,haggard_testing_2021}.
However, expectation values of general position-dependent observables cannot, in general, be obtained by simply evaluating the corresponding position-dependent function over the centroid probability distribution.
This is not to say that position dependent observables can not be inferred from the centroid distribution. 
~\citet{Kleinert1986} introduced a non-trivial expression that relates the expectation value of a position-dependent operator to a convolution of a centroid-based expectation value over the full configurational space. 
~\citet{cao_formulation_1994} subsequently generalized the relation to the exact centroid distribution in high dimensions.\\

A conceptual direction to simplify the calculation of expectation values of position-dependent observables—as explored in this work—is to construct quantum effective potentials that reproduce the exact position distribution defined by the thermal density matrix $(\hat{\rho}_{\beta}=e^{-\beta \hat{H}})$. 
This motivates the definition
\begin{equation}\label{eq:Veff_exact}
V_{\text{QM}}(x'; \beta) = -\beta^{-1} \log \rho_\beta (x') + \frac{1}{2\beta} \ln \frac{m}{2\pi\beta\hbar^2},
\end{equation}
for an effective potential, where $\rho_\beta (x') = \bra{x'} e^{-\beta\hat{H}}\ket{x'}$ denotes the diagonal element of the thermal density matrix in the position basis.
With this definition, the exact quantum partition function assumes a classical-like form,
\begin{align}
Z_{\textrm{QM}} = \sqrt{\frac{m}{2\pi \beta\hbar^2}}\int  \textrm{d}x'~ e^{-\beta V_{\text{QM}}(x'; \beta)},
\label{eq:classical_Z}
\end{align}
and the corresponding position distribution is
\begin{align}
P_{\text{QM}}(x') =\rho_{\beta}(x') ~/ ~Z_{\text{QM}}.
\end{align}
Expectation values of position-dependent observables then take the classical form
\begin{equation}
\left\langle O(\hat{x})\right\rangle = \int \mathrm{d}x'~O(x')~P_{\text{QM}}(x')
= \frac{\int \mathrm{d}x'~ O(x')~e^{-\beta V_{\text{QM}}(x';\beta)}}{\int \mathrm{d}x'~e^{-\beta V_{\text{QM}}(x';\beta)}}.
\label{eq:expectation_starting_point}
\end{equation}
On the one hand, despite the simplicity of Eq.~\ref{eq:expectation_starting_point}, effective potentials that approximate $V_{\text{QM}}(x';\beta)$ have received less attention than effective potentials based on the centroid coordinate.
On the other hand, local harmonic approximations have been used to approximate off-diagonal elements of the density operator in semiclassical dynamics (originally derived by \citet{ovchinnikov2001semiclassical} and later independently by \citet{shi2003semiclassical}, and also used by \citet{liu2009simple}).
This provides a natural starting point for exploring quantum effective potentials that approximate $V_{\text{QM}}(x';\beta)$, which is directly linked to the \textit{diagonal} element of the Boltzmann density matrix. \\



%
%
%
%

In this work, we propose a quantum effective potential by approximating the thermal density matrix within a local harmonic approximation.
In contrast to previous derivations for approximating off-diagonal elements, our approach does not require discretization of the imaginary-time path integral.
Instead, we treat quantum fluctuations in a mean-field manner about the starting point of the imaginary-time path, in the spirit of the Feynman--Hibbs and Feynman--Kleinert derivations, but without performing the mean-field treatment with respect to the path centroid.
We further introduce a series of  approximations for one-dimensional systems that enable extensions to non-convex potentials and yield excellent performance for strongly anharmonic systems.
Finally, we outline the potential usefulness of this approach in a broader computational chemical physics context.

 \section{Theory}

\subsection{The variational approach}

Within the path-integral formalism, an effective potential can be derived using the variational approach discussed by \citet{Feynman:100771}.
To elucidate this approach, we write the Euclidean action for a one-dimensional system in terms of the Lagrangian, parameterized by imaginary time $\tau \in [0,\beta\hbar]$
\begin{equation}\label{eq:euclidean_action}
    S \equiv S[x(\cdot); \beta\hbar] = \int_0^{\beta\hbar} \mathrm{d}\tau~ \left[\frac{1}{2}m\dot{x}(\tau)^2 + V\!\left(x(\tau)\right)\right].
\end{equation}
The relationship between the partition function, the Helmholtz free energy $F$ and action is
\begin{equation}\label{eq:helmholtz}
    Z_{\text{QM}} = e^{-\beta F} = \int \mathrm{d}x'~ \oint_{x'} \mathcal{D}[x(\cdot)]\, e^{-S[x(\cdot);\beta\hbar] / \hbar}.
\end{equation} \\

For any trial action $\tilde{S}$ and the corresponding free energy $\tilde{F}$, the following equality holds:
\begin{equation}
    e^{-\beta\left(F - \tilde{F}\right)} = \left\langle e^{-(S - \tilde{S}) / \hbar}\right\rangle_{\tilde{S}} .
\end{equation}
Here, $\langle \cdot \rangle_{\tilde{S}}$ denotes the expectation value of a functional over the space of closed-loop imaginary-time paths:
\begin{align}
    \langle f \rangle_{\tilde{S}} = \frac{\int dx' \oint_{x'} \mathcal{D} [x(\cdot)] \: f[x(\cdot)] \; e^{-{\tilde{S}[x(\cdot);\beta]}\slash{\hbar}}}{\int dx' \oint_{x'} \mathcal{D} [x(\cdot)] \:  e^{-{\tilde{S}[x(\cdot);\beta]}\slash{\hbar}} }.
    \label{eq:expectation_trial}
\end{align}
Noting that Eq.~\ref{eq:expectation_trial} induces a normalized measure proportional to $e^{-{\tilde{S}[x(\cdot);\beta]}\slash{\hbar}}$, applying Jensen's inequality, i.e.\ $\langle e^x \rangle \geq e^{\langle x \rangle}$, yields
\begin{align}
    F \leq \tilde{F} + (\beta\hbar)^{-1} \langle S - \tilde{S} \rangle_{\tilde{S}} .
    \label{eq:variational}
\end{align}
Since this inequality holds for any trial action functional $\tilde{S}$, it becomes tight when $S - \tilde{S}$ is path independent (in particular, for $\tilde{S}=S$). By selecting trial actions for which the required expectation values can be evaluated analytically, Eq.~\ref{eq:variational} provides a starting point for deriving quantum effective potentials.

\subsection{The centroid \textit{ansatz}}

A class of effective potentials can be derived in terms of the path centroid: $\bar{x} =  \frac{1}{\beta\hbar} \int_0^{\beta\hbar} \text{d}\tau~ x(\tau)$.
Eq.~\ref{eq:pi} can be reparametrized in terms of the centroid by defining the centroid-constrained functional integration measure as
$\mathcal{D}'[x(\cdot)] \equiv \mathcal{D}[x(\cdot)]\;
    \delta\!\left(\bar{x} - \frac{1}{\beta\hbar}\int_0^{\beta\hbar} x(\tau)\, d\tau\right)$. 
This yields the exact expression
\begin{align}
\label{eq:centroid_partition_function}
    Z_{\text{QM}} = \int \text{d}\bar{x} \int \text{d}x' \oint_{x'} \mathcal{D}'[x(\cdot)] e^{-{S[x(\cdot);\beta\hbar] /\hbar}} .
\end{align}
and an exact centroid effective potential $W(\bar{x}, \beta)$ that satisfies the relationship
\begin{align}
    e^{-\beta W(\bar{x})}
    = \sqrt{\frac{2\pi\beta\hbar^2}{m}}\,
    \int dx' \oint_{x'} \mathcal{D}'[x(\cdot)]\;
    e^{-{S[x(\cdot);\beta\hbar]/\hbar}}.
\end{align}
This defines the quantum partition function akin to the classical partition function
\begin{align}
    Z_{\text{QM}} = \sqrt{\frac{m}{2\pi\beta\hbar^2}} \int d\bar{x}\: e^{-\beta W(\bar{x})}.
\end{align}
While in practice, $W(\bar{x})$ can be evaluated from the exact constrained path integral, the variational approach can be used to analytical approximations to $W(\bar{x})$.

\subsubsection{The Feynman-Hibbs effective potential}

The Feynman--Hibbs effective potential is the first and simplest effective potential derived within the path-integral framework. 
It is obtained by decomposing the Euclidean action into that of a free particle and a potential contribution $S = S_{\text{free}} + S_{V}$ where the free particle action is defined as $S_{\text{free}} \equiv \int_0^{\beta\hbar} \text{d}\tau~ \left[\frac{1}{2}m\dot{x}(\tau)^2 \right]$.
The corresponding partition function of the free particle is $Z_{\text{free}} \equiv \int dx' \oint_{x'} \mathcal{D}[x(\cdot)]\, e^{-S_{\text{free}}/\hbar}.$
From Eq.~\ref{eq:expectation_trial}, the quantum partition function then takes the form
\begin{align}
    \label{eq:Z_decomp}
    Z_{\text{QM}} = Z_{\text{free}} \langle e^{-S_V/\hbar} \rangle_{S_{\text{free}}}.
\end{align} \\

The Feynman--Hibbs approximation follows by applying Jensen's inequality to Eq.~\ref{eq:Z_decomp}, which yields $\langle e^{-S_V/\hbar} \rangle_{S_{\text{free}}} \ge e^{-\langle S_V\rangle_{S_{\text{free}}}/\hbar}$, and motivates the replacement
\[\langle e^{-S_V/\hbar} \rangle_{S_{\text{free}}} \rightarrow e^{-\langle S_V/\hbar\rangle_{S_{\text{free}}}}\]
This substitution is followed by evaluating the free-particle expectation value by decomposing the path into its centroid and fluctuations and carrying out the resulting Gaussian functional integral over the fluctuations.
These non-trivial manipulations discussed in detail in Ref.~\citenum{Feynman:100771} lead to the following Gaussian integral for the resulting effective potential:
\begin{equation}
\label{eq:WFH}
V_{\text{FH}}\left(\bar{x}, \beta\right)
= \sqrt{\frac{6m}{\pi\beta\hbar^2}}
\int_{-\infty}^{\infty} \text{d}x~ e^{-\frac{6m}{\beta\hbar^2}{(x-\bar{x})^2}} ~V(x).
\end{equation}

\subsubsection{The Feynman-Kleinert effective potential}

Feynman and Kleinert improved on the effective potential in Eq.~\ref{eq:WFH} by introducing a more sophisticated trial action than $S_{\text{free}}$ and subsequently variationally minimising the corresponding trial free energy. 
Path fluctuations $(x(\tau)-\bar{x})$ are treated explicitly up to second order in the trial action and higher-order terms are absorbed into a term that depends only on the centroid.
This is motivated by expanding the potential about the path centroid
\begin{equation}\label{VTaylor_cent}
\begin{split}
V(x(\tau)) =& \: V(\bar{x}) + V'(\bar{x}) (x(\tau)-\bar{x}) \:+\\
& \frac{1}{2} V''(\bar{x}) (x(\tau)-\bar{x})^2 + \mathcal{O} (x(\tau)- \bar{x})^3 ,
\end{split}
\end{equation}
where $\mathcal{O} (x(\tau)- \bar{x})^3$ represents higher-order terms.
A harmonic trial action is then introduced,
\begin{equation}
\begin{split}
    \tilde{S}_{\text{FK}} \equiv & \int_0^{\beta\hbar} \text{d}\tau~ \left[\frac{1}{2}m\dot{x}(\tau)^2 \right.\\
    &+  \left. \frac{1}{2}V''(\bar{x}) \left(x(\tau) - \bar{x}\right)^2\right] - \beta\hbar L_1(\bar{x}).
\end{split}
\end{equation}
This action, like $S_{\text{free}}$, defines an analytically tractable reference measure while retaining the quadratic fluctuations about $\bar{x}$ explicitly.
$L_1(\bar{x})$ includes the zeroth-order contribution $V(\bar{x})$ and the mean-field effect of the higher-order terms in Eq.~\eqref{VTaylor_cent}. \\

Using $S_{\text{FK}}$ as the trial action in Eq.~\ref{eq:variational} yields a bound that depends on the correction term $\langle S - S_{\text{FK}}\rangle_{S_{\text{FK}}}$.
Under this measure defined by the trial action, fluctuations about the centroid are symmetric, so $\langle x(\tau)-\bar{x}\rangle_{S_{\text{FK}}}=0$.  
The linear term in Eq.~\eqref{VTaylor_cent} also does not contribute to $\langle S - S_{\text{FK}}\rangle_{S_{\text{FK}}}$.
Because $S_{\text{FK}}$ defines a Gaussian measure for the fluctuations, the remaining contributions reduce to Gaussian functional averages, which can be evaluated analytically.
It is convenient to parameterize these statistics by the fluctuation amplitude $a^2$ and an effective fluctuation frequency $\Omega$; these quantities are initialized by the harmonic reference and are optimized using the by variational minimisation.
These steps yield the Feynman--Kleinert effective potential
\begin{equation}\label{eq:WFK}
\begin{split}
    V_{\text{FK}}(\bar{x}, \beta;a^2,\Omega) & = \frac{1}{\beta\hbar}\ln \frac{\sinh (\beta\hbar\Omega/2) }{\beta\hbar\Omega/2} \\ & - \frac{m\Omega^2}{2}a^2 + V_{a^2}(\bar{x}, \beta),
\end{split}
\end{equation}
where $V_{a^2}(\bar{x})$ is the smeared potential
\begin{align}\label{Va2}
    V_{a^2}(\bar{x}, \beta) = \frac{1}{\sqrt{(2\pi a^2)}} \int_{-\infty}^{\infty} \text{d}x~  e^{-\frac{(\bar{x}-x)^2}{2a^2} } V(x) .
\end{align}
We refer the reader to \cite{FeynmanKleinert} for more details on the derivation and implementation of this effective potential.

\subsection{The starting point \textit{ansatz}}

Centroid-based effective potentials are useful for estimating partition functions of low-dimensional systems~\cite{Feynman:100771, FeynmanKleinert} and for obtaining approximate dynamical information from the centroid potential of mean force~\cite{cao_new_1993, cao_formulation_1994}.
Often, expectation values of nonlinear position-dependent operators $O(\hat{x})$ are of interest.
Such expectation values can be obtained from the partition function by coupling $O$ to the potential and differentiating, i.e.
\begin{equation}
\langle O(\hat{x}) \rangle
= -\frac{1}{\beta}\left.\frac{\partial}{\partial \lambda}\ln Z_{V(x)\to V(x)+\lambda O(x)}\right|_{\lambda=0}.
\end{equation}
As is evident from Eq.~\ref{eq:centroid_partition_function} and its derivation, the centroid formulation marginalizes over the centroid coordinate rather than the physical position, so a bespoke centroid-based estimator must be derived for each observable $O$.
This complication was noted in \citet{Kleinert1986}, a follow-up to the original Feynman--Kleinert potential~\cite{FeynmanKleinert}, where a non-trivial expression was introduced that relates the expectation value of a position-dependent operator to a convolution of a centroid-based expectation value over the full configurational space. \\

To render simplicity to to the evaluation of position dependent observables, we explore quantum effective potentials parametric to the \textit{starting point} $x' \equiv x(0)$ of the Feynman paths rather than the path. 
%
%
The motivation for this ansatz, as illustrated in the derivation in  Appendix~\ref{appendix:starting_point_estimator}, is that the expectation value of a position-dependent operator depends only on its value at the path starting point $x'$, even though the action depends on the full path, i.e.
\begin{equation}
\begin{split}
    \langle O(\hat{x}) \rangle & =  \frac{\int \text{d}x' ~O(x')~\oint_{x'} \mathcal{D}[x(\cdot)]\, e^{-\hbar^{-1} S[x(\cdot);\beta\hbar]}}{\int \text{d}x' ~\oint_{x'} \mathcal{D}[x(\cdot)]\, e^{-\hbar^{-1} S[x(\cdot);\beta\hbar]}}.
\end{split}
\end{equation}
Another way of interpreting this equation is that the expectation value of a position-dependent observable can be estimated as an ensemble average of $O(x')$ over the distribution of path starting points, motivating the derivation of analytic potentials that target this distribution.

\subsubsection{Local harmonic approximation}

As in the Feynman–Kleinert construction, we employ a quadratic trial action; however, we formulate it by expanding about the path starting point $x'$ rather than the centroid.
We expand the potential along the path as
\begin{equation}
\label{eq:VTaylor_sp}
\begin{split}
V(x(\tau)) =& \: V(x') + V'(x') \bigl(x(\tau)-x'\bigr) \:+\\
& \frac{1}{2} V''(x') \bigl(x(\tau)-x'\bigr)^2 + \mathcal{O}\!\bigl(x(\tau)- x'\bigr)^3 ,
\end{split}
\end{equation}
where $\mathcal{O}\!\bigl(x(\tau)- x'\bigr)^3$ denotes higher-order terms about $x'$.
Truncating Eq.~\eqref{eq:VTaylor_sp} at second order motivates the local harmonic trial action
\begin{equation}
\begin{split}\label{eq:SLH}
    S_{\text{LH}} \equiv & \int_0^{\beta\hbar} \mathrm{d}\tau~ \left[\frac{1}{2}m\dot{x}(\tau)^2 +  V'(x)~(x(\tau)-x'\bigr) \right.\\
    &+  \left. \frac{1}{2}V''(x)~(x(\tau)-x')^2\right] + \beta\hbar V(x').
\end{split}
\end{equation}
To ensure that $S_{\text{LH}}$ defines a well-behaved harmonic reference measure, we assume $V''(x')>0$ unless stated otherwise.
Using $S_{\text{LH}}$ as the trial action in Eq.~\ref{eq:variational}, the deviation from the exact free energy is controlled by
\begin{align}
\langle S - S_{\text{LH}}\rangle_{S_{\text{LH}}}
= \left\langle \int_{0}^{\beta\hbar} \text{d}\tau~ \mathcal{O} (x(\tau)-x')^3 \right\rangle_{S_{\text{LH}}},
\end{align}
which suggests that the difference between the exact and local-harmonic free energies arises entirely from the higher-order terms in Eq.~\eqref{eq:VTaylor_sp}. For a purely harmonic potential the remainder vanishes and $S=S_{\text{LH}}$, so the approximation becomes exact. \\

In what follows, we construct an effective potential directly from the local-harmonic trial action, without variationally minimizing the deviation between the true and trial actions.
We approximate the diagonal density matrix element as
\begin{align}
    \rho_{\beta}(x') \approx \rho^{\text{LH}}_{\beta}(x') = \oint_{x'} \mathcal{D}[x(\cdot)]\, e^{-{\hbar^{-1} S_{\text{LH}}[x(\cdot);\beta\hbar]}} .
\end{align}
The next step is to rewrite $S_{\text{LH}}$ by completing the square in the displacement $x(\tau)-x'$,
\begin{equation}\label{eq:SLH_csq}
\begin{split}
S_{\text{LH}} =& \frac{m}{2} \int_0^{\beta\hbar}\mathrm{d}\tau~ \left[\dot{x}(\tau)^2  + \omega_a^2 \bigl( x(\tau)- x_0 \bigr )^2 \right] \\ &
    - \beta\hbar\frac{m k_a^2}{2\omega_a^2} +\beta \hbar V(x'),
\end{split}
\end{equation}
where we introduce the shorthands for the origin, instantaneous acceleration, and instantaneous angular frequency of the locally harmonic component, respectively:
\[x_0 \equiv x' - \frac{k_a}{\omega_a^2},\quad
k_a \equiv \frac{1}{m}\left.\frac{\partial V}{\partial x}\right|_{x=x'},\quad
\omega_a^2 \equiv \frac{1}{m}\left.\frac{\partial^2 V}{\partial x^2}\right|_{x=x'}\].

Equation~\eqref{eq:SLH_csq} expresses the full action as the sum of the Euclidean action of a harmonic oscillator with frequency $\omega_a$ and equilibrium position $x_0$, which depends quadratically (explicitly) on the path, plus terms that are independent of the path.
Consequently, $\rho^{\mathrm{LH}}_{\beta}(x')$ factorizes into the diagonal thermal density matrix element of a harmonic oscillator with $(x_{\mathrm{eq}},\omega)=(x_0,\omega_a)$ evaluated at $x'$ (for which an analytic expression exists) and an exponential prefactor arising from the path-independent terms, yielding an overall closed-form expression:
\begin{equation}\label{eq:rho_LH}
\begin{split}
 \rho^{\text{LH}}_{\beta}(x') =&\sqrt{\frac{m}{2\pi\beta\hbar^2}}\sqrt{\frac{2\xi_a}{\sinh 2\xi_a}}  \times \\ &
\exp\left\{-\beta \left [V(x') + \frac{m k_a^2}{2\omega_a^2} \left( \frac{\tanh \xi_a}{\xi_a} - 1 \right)\right]\right\},
\end{split}
\end{equation}
where a shorthand $\xi_a \equiv \frac{\beta\hbar\omega_a}{2}$ is introduced for the temperature scaled instantaneous harmonic frequency.
The corresponding locally harmonic effective potential derived from the starting-point ansatz can be obtained by substituting Eq.~\ref{eq:rho_LH}
\begin{equation}
\label{eq:Veff_prelim}
\begin{split}
\tilde{V}_{\text{LH}}(x') &= V(x')  + \frac{m k_a^2}{2\omega_a^2}  \left(\frac{\tanh \xi_a}{\xi_a} -1\right) \\ & + \frac{1}{2\beta}\ln \frac{\sinh 2\xi_a}{2\xi_a}.
\end{split}
\end{equation} 

We note that \citet{ovchinnikov2001semiclassical} applied a local harmonic approximation to a more general quantity, namely the off-diagonal matrix element of the Boltzmann operator $\mel{x'}{e^{-\beta \hat{H}}}{x''}$.
In contrast to our derivation, they obtain an expression for this off-diagonal element by applying a local harmonic approximation within a discretized functional integral, i.e., by expanding the potential energy for each path integral ``replica" in a Taylor series about the path starting point and truncating at second order.
Eq.~\ref{eq:Veff_prelim} can be recovered by setting $x-x'=0$ in their expression, which can be seen as a validation of our derivation.
In addition, our formulation naturally suggests a route to a more flexible variant of Eq.~\ref{eq:Veff_prelim} for a given system by treating $x_0$, $k_a$, and $\omega_a$ as variational parameters.

\subsubsection{Further approximations}\label{Subsec:Renorm}

This expression yields the exact partition function in the classical (i.e. $\beta\rightarrow 0$) and harmonic limits (see Appendix \ref{appendix:harm_cl_limit}). However, in anharmonic potentials where the curvature $\omega_a^2$ can take negative values, \eqref{eq:rho_LH} cannot be used to obtain the effective potential, as it is derived under the assumption that $\omega_a$ is real. Physically, this is because tunneling dominates quantum statistics around points with a negative curvature and a second-order truncation of the Taylor expansion of the potential $V$ (which assumes a compact imaginary-time Feynman path) is no longer valid. One way around this issue is to force $\omega_a^2 \geq 0$ to ensure the well-definedness of the effective potential, at the cost of neglecting tunneling contributions. However, even with such frequency cutoffs, the second and third (non-classical) terms in \eqref{eq:Veff_prelim} can potentially diverge around regions where $\omega_a^2 \rightarrow 0$. It is straightforward to see that the second term is always negative when $\omega_a^2>0$ and it becomes more unstable when $\beta \rightarrow \infty$ and $\omega_a^2 \rightarrow 0 $. The third term approaches the limit $\omega_a/2$ as $\beta \rightarrow \infty$ at any given point $x'$. When $\omega_a \rightarrow 0$, this term also causes a sharp (and potentially unphysical) localisation around $x'$. We hence refine this potential further for it to be computationally useful. \\

The effective potentials in previous sections (given by Eqs.~\eqref{eq:WFH}, \eqref{eq:WFK} and \eqref{eq:Veff_prelim}) were derived with the primary aim of approximating the quantum \textit{partition function}.
However, for statistical averages of position-dependent observables, it is sufficient to approximate the quantum position distribution $P_{\text{QM}}(x')$, obtained by normalizing the diagonal density-matrix elements by the quantum partition function.
While $Z_{\text{QM}}\to 0$ as $\beta\rightarrow\infty$, the normalized distribution $P_{\text{QM}}(x')$ remains well-defined due to zero-point fluctuations.
The exact quantum PDF is therefore
\begin{align}
    P_{\text{QM}}(x') = \rho_{\beta}(x') ~/~Z_{\text{QM}}.
\end{align}
Motivated by this, we introduce a renormalized local-harmonic approximation that targets $P_{\text{QM}}(x')$ directly.
We distinguish the unrenormalized local-harmonic diagonal density matrix, $\rho^{\text{LH}}_{\beta}(x')$, from a renormalized variant, $\tilde{\rho}^{\:\text{LH}}_{\beta}(x')$, defined as
\begin{align}
\label{eq:renorm}
    \tilde{\rho}^{\:\text{LH}}_{\beta}(x') = \,\rho^{\text{LH}}_{\beta}(x') ~/~ \Upsilon_{\text{NCF}}(x'),
\end{align}
where $\Upsilon_{\text{NCF}}(x')$ is a local renormalisation factor,
\begin{align}\label{ZNCF}
    \Upsilon_{\text{NCF}}(x') = Z_{\text{QM,local}}(x') ~/~ Z_{\text{cl,local}}(x').
\end{align}
Here $Z_{\text{cl,local}}(x') = 1/(2\xi_a)$ and $Z_{\text{QM,local}}(x') = 1/(2\sinh \xi_a)$ are the \textit{local} classical and quantum partition functions of a notional harmonic oscillator with frequency $\omega_a$.
By construction, $\Upsilon_{\text{NCF}}(x')\to 1$ in the classical limit, and becomes independent of $x'$ in the harmonic limit.
Finally, an approximate normalized position distribution is obtained by globally normalizing $\tilde{\rho}^{\:\text{LH}}_{\beta}(x')$,
\begin{align}
    P^{(r)}_{\text{LH}}(x') := \frac{\tilde{\rho}^{\:\text{LH}}_{\beta}(x')}{\int \mathrm{d}x''~\tilde{\rho}^{\:\text{LH}}_{\beta}(x'')}.
\end{align}

This scaling prevents artifacts in the resulting position density and removes unphysical divergences in the $\beta\rightarrow\infty$ limit.
With this renormalisation, the local-harmonic reweighted density takes the form
\begin{equation}\label{Prenorm}
\begin{split}
    {P}^{(r)}_{\text{LH}}(x') 
    =&\sqrt{\frac{m}{2 \pi\beta\hbar^2}} \sqrt{\frac{\tanh \xi_a}{\xi_a}}  \exp \left \{-\beta \left (V(x') - \frac{m k_a^2}{2\omega_a^2}  \right) \right \} \\&\times\exp\left\{- \beta\frac{m k_a^2}{2\omega_a^2} \left( \frac{\tanh \xi_a}{\xi_a} \right)\right\} .
\end{split}
\end{equation}

However, even this form doesn't address all computational issues: while ($V(x') - \frac{m k_a^2}{2\omega_a^2}$) vanishes identically for a harmonic oscillator, in the limit $\beta \rightarrow\infty$, its presence in the first exponential yields a $\mathcal{O}(\beta)$ leading order term in the effective potential for generic anharmonic systems. 
This $\beta$-dependent drift could be reduced by treating $\omega_a$ and $x_0$ as variational parameters; here, we pursue simpler approximate expressions.
To this extent, we apply the following \textit{harmonic mapping substitution} in Eq.~\eqref{Prenorm},
\begin{align}\label{magic}
    \frac{m k_a^2}{2\omega_a^2} \rightarrow V(x') ,
\end{align}
and introduce the non-classical smearing factor $\Xi(x') = \frac{\tanh \xi_a}{\xi_a}$, which yields
\begin{align}\label{PLH_final}
    {P}_{\text{LH}}(x')  =\sqrt{\frac{m}{2 \pi\beta\hbar^2}} \sqrt{\Xi(x')} \exp{-\beta V(x') \Xi(x') } ,
\end{align}
To obtain a normalized distribution, we define
\begin{align}\label{PLH_norm}
    \overline{P}_{\text{LH}}(x') := \frac{P_{\text{LH}}(x')}{\int \mathrm{d}s~P_{\text{LH}}(s)} .
\end{align}
The effective classical potential is then obtained by taking the logarithm of Eq.~\eqref{PLH_final},
\begin{align}\label{VLH_final}
    V_{\text{LH}}(x') = V(x')\Xi(x') - \frac{1}{2\beta} \ln \Xi(x') .
\end{align}
Equations \eqref{PLH_final} and \eqref{VLH_final} are the main results of this article and will be referred to as the local-harmonic distribution function and local-harmonic effective classical potential henceforth.





\section{Computational details}

To assess the accuracy with which quantum effective potentials reproduce position-dependent operators, we examine their ability to reproduce the position distribution function. 
To obtain the quantum reference, we compute the numerically exact position distribution function by exact diagonalization in a discrete position basis, implemented in Python using the standard NumPy library.
For a given configurational space the results are converged with respect to the discretization of the position grid. 
The Feynman-Hibbs and Feynman-Kleinert effective potentials for polynomial potentials  ($V(x)=x^n$) can be calculated by recognising that the smeared potential $V_{a^2}$ in Eqs.~\ref{eq:WFH} and \ref{eq:WFK} have an exact analytical form as shown in Appendix \ref{appendix:FH and FK evaluation}. 
Note that $a^2 = \beta \hbar^2/3m$ for the Feynman-Hibbs effective potential \eqref{eq:WFH}, whereas $a^2$ is treated as a variational parameter in the Feynman--Kleinert effective potential alongside $\Omega$ in \eqref{eq:WFK}. 

The non-linear equations
\begin{align}
    a^2(\bar{x}) &=  \frac{1}{\beta\Omega^2(\bar{x})} \left[ \frac{\beta\Omega(\bar{x})}{2} \coth\frac{\beta\Omega(\bar{x})}{2} -1  \right] \\
    \Omega^2(\bar{x}) &= \frac{\partial}{\partial x^2_0} V_{a^2} (\bar{x}),
\end{align}
are solved self-consistently to obtain the optimal values. For potentials which are not polynomials, one needs to evaluate the smeared potential $V_{a^2}$ using numerical integration.

%

\section{Numerical results}

\subsection{Harmonic and Perturbed Harmonic Potential}

We demonstrate the accuracy of the approximation for generic anharmonic systems of the form (as used by Ref.~\citet{FeynmanKleinert} to estimate the quantum partition functions):
\begin{align} \label{Vquartic}
    V(x) = \frac{1}{2}m\omega^2x^2 + \frac{g}{4} x^4
\end{align} with $m=1, \omega=1$ in Fig.~\ref{fig1a}-\ref{fig1d}.
Here, we use this potential to assess the accuracy of the position distribution function by comparison against the numerically exact distribution at three different inverse temperatures: $\beta \in \{0.1, 1.0, 10.0\}$.
We also include comparisons with the classical position distribution function, as well as the classical distributions obtained using the Feynman--Hibbs and Feynman--Kleinert effective potentials.
Since we do not compute the rigorous position distribution functions corresponding to the Feynman--Hibbs and Feynman--Kleinert effective potentials, due to the cumbersome associated equations, comparisons with the reference should therefore be treated with caution.
Accordingly, deviations of these centroid distributions should not be interpreted as errors in the underlying approximations to the rigorous position distribution function, but rather as differences between centroid distributions and the rigorous position distribution. \\

\begin{figure*}
\centering
\includegraphics[width=\textwidth]{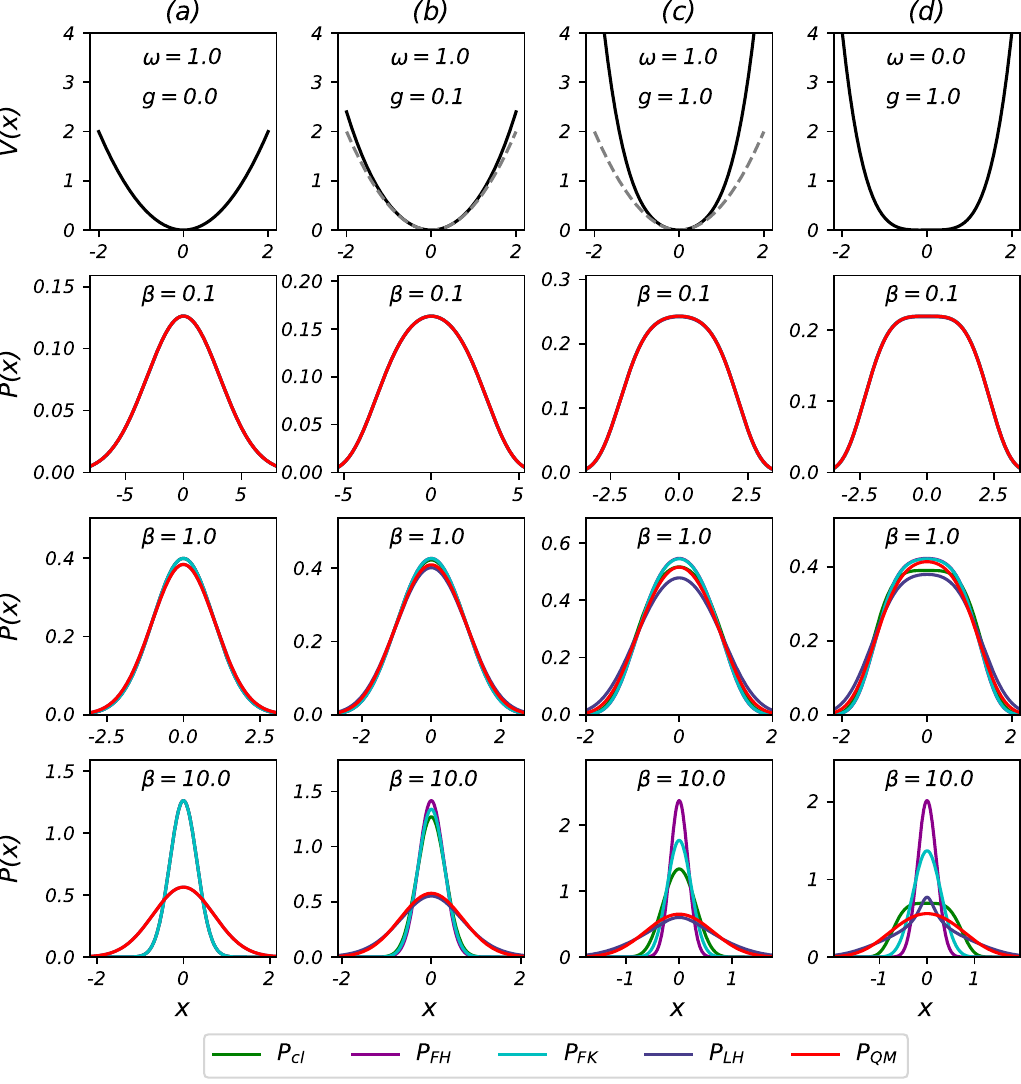}
\caption{Comparison of the classical (green), quantum (red),
Feynman-Hibbs (purple), Feynman-Kleinert (cyan)
and local harmonic (dark blue) position distribution functions (from \eqref{PLH_final}) at different temperatures for the quartic potential potential $V(x) = m\omega^2x^2/2 + gx^4/4$  in \eqref{Vquartic} at different values of quartic perturbation strength $g$. The top row contains plots of the respective potential energy surfaces along with the reference harmonic oscillator of frequency $\omega=1.0$ (indicated in grey dashed line).  Note that, for the purely harmonic oscillator with $g=0.0$, the Feynman-Kleinert
and Feynman-Hibbs PDFs are hidden behind the classical PDF and the local harmonic PDF is hidden behind the quantum PDF.}
\labelphantom{fig1a}
\labelphantom{fig1b}%
\labelphantom{fig1c}%
\labelphantom{fig1d}
\end{figure*}

When $g=0$, Eq.~\eqref{Vquartic} reduces to the harmonic oscillator.
In this limit, the exact results for all distributions are known analytically; it therefore serves as a numerical validation of our implementation and a reference point for the anharmonic regimes of Eq.~\eqref{Vquartic}.
The exact distribution can be obtained analytically, and the distribution from Eq.~\eqref{PLH_final} can be shown analytically to reproduce the exact harmonic-oscillator result, as seen in the figure.
Furthermore, while the Feynman--Kleinert and Feynman--Hibbs methods predict the partition function correctly for the harmonic oscillator, they yield a classical centroid distribution, since the quantum corrections in these effective potentials enter the prefactor of the partition function rather than the position dependence; this is also evident in the figure. \\

We next introduce mild and strong quartic perturbations to the harmonic oscillator by setting $g=0.1$ and $g=1.0$, respectively, in Fig.~\ref{fig1d} and Fig.~\ref{fig1d}.
In the small-perturbation regime, the Feynman--Kleinert and Feynman--Hibbs centroid distributions remain close to the classical reference; however, they err on the side of being more localized than the classical distribution.
This direction of error is more pronounced in the large-perturbation limit, with the Feynman--Kleinert approach yielding a centroid distribution that is closer to the quantum distribution than Feynman--Hibbs.
Finally, we assess the accuracy of the harmonically mapped local harmonic approximation.
We find near-quantitative agreement with the reference, with a residual error corresponding to a slight over-correction relative to the classical distribution.
The good agreement of the harmonically mapped local harmonic approximation is not a matter of chance.
Instead, it comes from careful approximations introduced in Sec.~\ref{Subsec:Renorm}, namely the renormalization in Eq.~\eqref{eq:renorm} and the harmonic mapping in Eq.~\eqref{magic}.
We analyse the impact of these approximations in more detail in the supporting information.
As shown in Fig.~S1 of the SI, the raw local harmonic approximation yields a distribution that errs on the classical side.
This arises from the form of the smearing potential in Eq.~\eqref{eq:WFH}, which biases the resulting distribution towards being more localized around the origin, where the curvature is smallest.
The \textit{non-classical} renormalization in Eq.~\eqref{eq:renorm} yields a less localized distribution (at the expense of an additive offset in the potential, which does not affect the position distribution). 
Finally, the harmonic mapping alleviates the explicit temperature dependence of this localization. \\

The most stringent test for effective potentials is the quartic oscillator with $\omega=0$ (and $g=1.0$), which lacks any harmonic support.
We show the position distribution functions from different approaches in Fig.~\ref{fig1d}.
The Feynman--Kleinert, Feynman--Hibbs, and the `bare' local harmonic effective potential in Eq.~\eqref{Veff_prelim} yield distributions that are strongly peaked around the origin, even relative to the classical distribution.
The harmonically mapped local harmonic approximation yields a distribution with a classical-like localization around the origin, while the overall width is in good agreement with the exact quantum result.
Based on these experiments, we conclude that our proposed approximation is best suited for potentials with a harmonic component.

\begin{figure}
    \centering
    \includegraphics[width=\linewidth]{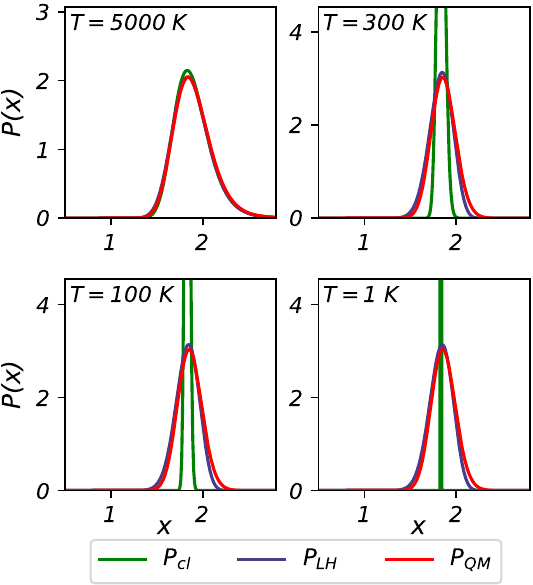}
    \caption{ Comparison of the classical (green), quantum (red) and local harmonic (dark blue) position distribution functions (from \eqref{PLH_final}) for the morse potential \eqref{Morsepot} with parameters as in \cite{rossi2014remove} at different temperatures. The dotted black lines correspond to the `bare' PDF obtained from \eqref{Veff_prelim} without the transformation in \eqref{magic} (For T=1K, the unphysical divergence of the effective potential \eqref{Veff_prelim} was too serious to result in a PDF). }
    \label{fig:morse_trpmd}
\end{figure}

\subsection{Morse potential and the double well potentials}

We next consider two physical models relevant to chemistry, starting with the Morse oscillator.  
This system is challenging as it exhibits a negative curvature in parts of the potential energy function.
We consider a one-dimensional Morse potential of the form:
\begin{align}\label{Morsepot}
    V(x) = D (1-e^{-\alpha (x-x_e)})^2
\end{align}
with the potential parameters $D$ and $\alpha$ given by the vibrational potential energy of the OH-bond  (as in Ref.~\cite{rossi2014remove}):
\[D = \frac{hc \:\omega_e^2}{4\omega_e \chi_e}, \qquad
    \alpha = \sqrt{\frac{2\mu_{\text{OH}} h c \:\omega_e \chi_e}{\hbar^2}}\]   
with $\omega_e = 3737.76 \:\text{cm}^{-1}$, $\omega_e \chi_e = 84.881 \: \text{cm}^{-1}$, $x_e = 0.9697 \AA$ and $\mu_{\text{OH}}$ is the reduced mass of the OH bond. 
The classical, quantum and local harmonic position distribution functions at four different temperatures (corresponding to four different temperature regimes) are plotted in Fig.~\ref{fig:morse_trpmd}. 
It can be observed that the harmonic mapped local harmonic ansatz for the distribution function proves near quantitatively accurate at all temperatures considered. 
In the SI, we show that the bare distribution function obtained from Eq.\eqref{Veff_prelim} becomes sharply localised around the region with a negative curvature which renders it inaccurate. 
This demonstrates the usefulness of the harmonic mapping substitution \eqref{magic}. \\

\begin{figure}
    \centering
    \includegraphics[width=\linewidth]{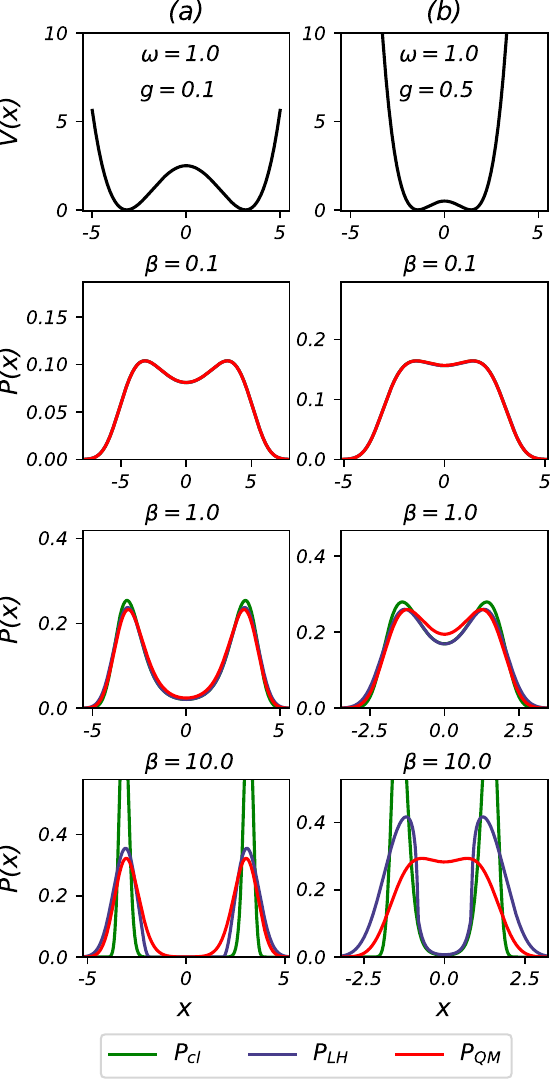}
    \caption{Same as in fig.~\ref{fig:morse_trpmd} but for the double-well potential $-m\omega^2x^2/2 + gx^4/4 + {m\omega^4}/ {16g}$ as in \eqref{DWPES} with (a) $g=0.1$ and (b) $g=0.5$ at different temperatures. The top row displays the potential energy surface indicating the well depths for both these parameter regimes. }
    \labelphantom{fig3a}
    \labelphantom{fig3b}
\end{figure}

Finally, we consider the double well potential, a simple physical for bond breaking and quantum tunnelling probabilities. 
We use a similar functional form as in \eqref{Vquartic}:
\begin{align}\label{DWPES}
    V(x) = -\frac{1}{2}m\omega^2x^2 + \frac{g}{4}x^4 + \frac{m\omega^4}{16g}
\end{align}
with $\omega = 1.0, m=1.0$. 
We consider two different values of the parameter $g$: $g=0.1$ and $g=0.5$. 
The main difference between these two values is the depth of the wells; 
$g=0.1$ corresponds to a deep well
and is consequently less anharmonic (i.e. with a large density of states inside the wells). 
$g=0.5$ yields a rather shallow double well which is very anharmonic. 
A comparison of the different position distribution functions in three different inverse temperature regimes $\beta \in \{0.1, 1.0, 10\}$ is shown in fig.~\ref{fig3a} and fig.~\ref{fig3b} for $g=0.1$ and $g=0.5$ respectively. \\

The harmonically mapped local harmonic position distribution function in \eqref{PLH_final} proves displays semi-quantitative accuracy with respect to the localization of the minima for $g=0.1$. 
As with the Morse oscillator, the bare local harmonic position distribution function yields an unphysical localisation around the region where the curvature of the potential changes sign. 
However, both the harmonically mapped and bare local harmonic position distribution functions fail to localize the minima in the shallow double well with $g=0.5$ in the lowest temperature regime considered.
In this parameter regime, the number of states within the wells is less than or equal to one, making the quantum effects from tunneling very prominent. 

\section{Discussion and Conclusions}

In summary, we have derived expressions of quantum effective potentials that approximate the potential of mean force associated with the thermal density distributions.
Our derivation closely follows those by variational approach suggested by Feynman and Hibbs~\cite{Feynman:100771}but differs in that we mean-field the quantum fluctuations about the path starting point rather than the path centroid. 
Instead of introducing parameters and variationally optimizing the potential, we explore a set of  approximation to render the potential numerically robust for general anharmonic potentials exhibiting negative curvatures. 
Apart from improved numerical stability, we do not have a clear justification for these approximations, other than that they become exact in the harmonic limit.
Deriving an effective potential parametric to the path starting point offers a benefit over centroid based effective potentials in that position dependent observables can be obtained as simple ensemble averages of the corresponding functions. 
Using numerical computation of the position distribution functions across a range of temperatures, we showed that this method is indeed effective (quantitatively accurate) in simulating statistical properties of a range of model potentials that have a strong harmonic support, including chemical model systems like a morse potential or a separated double well potential. 
We note quantitative errors for potentials such as a quartic potential and a shallow double well potential. 
We are optimistic that some of these limitations may be attenuated by variationally optimizing some of the parameters of the parameters in the potential, in the spirit of the Feynman-Kleinert effective potential~\cite{FeynmanKleinert}. \\
While the one-dimensional systems considered here help clarify the strengths and limitations of the derived effective potentials, they do not correspond to the settings in which inexpensive yet accurate approximations to thermal density distributions are most needed.
Looking ahead, we see two ways in which the present work may be useful in this broader context.
First, it provides the groundwork for deriving analogous expressions in higher dimensions; both the resulting analytic forms and simulations beyond the harmonic approximation, at a computational overhead comparable to classical methods, could be useful, for example, in the lattice dynamics of anharmonic solids~\cite{Werthamer1970}.
Second, our results may help motivate functional forms or constraints for training machine-learning models of thermal density distributions.
While machine-learning models have frequently been used to represent the centroid potential of mean force~\cite{musil_physics-inspired_2021, loose_centroid_2022}, learning the potential of mean force associated with the path starting point directly from path-integral molecular dynamics has proven more challenging~\cite{zaporozhets_accurate_2024}.
We hope that the expressions derived here can motivate physics-based architectures for this problem, in the same way that incorporating physical inductive biases into model design has benefited the machine-learning interatomic potentials community~\cite{behler_generalized_2007, bartok_gaussian_2010, grisafi_incorporating_2019}.

\section*{Acknowledgement}
This work was funded by the Ernest Oppenheimer Fund (through an early career fellowship) and the Leverhulme Trust. V.G.S also acknowledges funding from the Max Planck Institute for the Physics of Complex Systems, Dresden.

\section*{Data Availability}
The data will be made available upon publication.

\appendix

\appendix
\section{Starting-point estimator for position-dependent observables}
\label{appendix:starting_point_estimator}

We derive an estimator for $\langle O(\hat{x})\rangle$ for position-dependent observables $O(\hat{x})$ by considering the weakly perturbed Hamiltonian
\begin{equation}
\hat{H}(\lambda)=\frac{\hat{p}^2}{2m}+V(\hat{x})+\lambda O(\hat{x}),
\end{equation}
with $\lambda\in\mathbb{R}$.
The corresponding partition function is
\begin{equation}
Z(\lambda)=\mathrm{Tr}\!\left[e^{-\beta \hat{H}(\lambda)}\right]
=\int \mathrm{d}x'~\mel{x'}{e^{-\beta \hat{H}(\lambda)}}{x'}.
\end{equation}
Differentiating and using cyclicity of the trace yields
\begin{equation}
\frac{\partial}{\partial\lambda}\ln Z(\lambda)
=\frac{1}{Z(\lambda)}\frac{\partial Z(\lambda)}{\partial\lambda}
=-\beta\,\frac{\mathrm{Tr}\!\left[O(\hat{x})e^{-\beta \hat{H}(\lambda)}\right]}{\mathrm{Tr}\!\left[e^{-\beta \hat{H}(\lambda)}\right]}.
\end{equation}
Evaluating at $\lambda=0$ gives the standard thermodynamic relation
\begin{equation}
\langle O(\hat{x})\rangle
= -\frac{1}{\beta}\left.\frac{\partial}{\partial\lambda}\ln Z(\lambda)\right|_{\lambda=0}.
\label{eq:app_coupling_identity}
\end{equation}

To obtain a starting-point expression, we write the numerator of Eq.~\eqref{eq:app_coupling_identity} in the position basis:
\begin{equation}
\mathrm{Tr}\!\left[O(\hat{x})e^{-\beta\hat{H}}\right]
=\int \mathrm{d}x'~\mel{x'}{O(\hat{x})e^{-\beta\hat{H}}}{x'}.
\end{equation}
Since $O(\hat{x})\ket{x'}=O(x')\ket{x'}$, the matrix element factorizes as
\begin{equation}
\mel{x'}{O(\hat{x})e^{-\beta\hat{H}}}{x'}
=O(x')\,\mel{x'}{e^{-\beta\hat{H}}}{x'}.
\label{eq:app_factorize}
\end{equation}
Substituting Eq.~\eqref{eq:app_factorize} into Eq.~\eqref{eq:app_coupling_identity} and using
$Z_{\text{QM}}=\int \mathrm{d}x'~\mel{x'}{e^{-\beta\hat{H}}}{x'}$
yields
\begin{equation}
\langle O(\hat{x})\rangle
=\frac{\int \mathrm{d}x'~O(x')\,\mel{x'}{e^{-\beta\hat{H}}}{x'}}{\int \mathrm{d}x'~\mel{x'}{e^{-\beta\hat{H}}}{x'}}.
\label{eq:app_starting_point_dirac}
\end{equation}

Finally, we express the diagonal imaginary-time propagator as a closed-path integral with basepoint $x'$,
\begin{equation}
\mel{x'}{e^{-\beta\hat{H}}}{x'}
=\oint_{x'} \mathcal{D}[x(\cdot)]\,e^{-\hbar^{-1} S[x(\cdot);\beta\hbar]},
\end{equation}
and substitute into Eq.~\eqref{eq:app_starting_point_dirac} to obtain the path-integral estimator
\begin{equation}
\langle O(\hat{x}) \rangle
=  \frac{\int \mathrm{d}x' ~O(x')~\oint_{x'} \mathcal{D}[x(\cdot)]\, e^{-\hbar^{-1} S[x(\cdot);\beta\hbar]}}{\int \mathrm{d}x' ~\oint_{x'} \mathcal{D}[x(\cdot)]\, e^{-\hbar^{-1} S[x(\cdot);\beta\hbar]}}.
\label{eq:app_starting_point_path}
\end{equation}

\section{Classical and Harmonic Limit of the Effective Potential}\label{appendix:harm_cl_limit}
We start with the `bare' effective potential and PDF as in \eqref{Veff_prelim}:
\begin{dmath}\label{Veff_prelim}
   \tilde{V}_{\text{eff}}(x_a) = V(x_a)  +
     \frac{m k_a^2}{2\omega_a^2}  \left(\frac{\tanh \xi_a}{\xi_a} -1\right) + \frac{1}{2\beta}\ln \frac{\sinh 2\xi_a}{2\xi_a}
\end{dmath}
As $\beta \rightarrow 0$ (high-temperature/classical limit), we have:
\begin{align}
   \frac{\tanh \xi_a}{\xi_a} \rightarrow 1 \quad\quad \frac{\sinh 2\xi_a}{2\xi_a} \rightarrow 1
\end{align}
which gives
\begin{align}
    \tilde{V}_{\text{eff}}(x_a) \rightarrow V(x_a)
\end{align}
For a harmonic oscillator $V(x) = m\omega^2x^2/2$, we get:
\begin{align}
    \frac{mk_a^2}{2\omega_a^2} = \frac{1}{2} m \omega^2 x_a^2
\end{align}
which simplifies the effective potential \eqref{Veff_prelim} to:
\begin{align}
     \tilde{V}_{\text{eff}}(x_a) = \frac{m\omega x^2}{\hbar} \tanh \left(\frac{\beta \hbar \omega}{2}\right)+\frac{1}{2\beta}\ln \frac{\sinh \beta\hbar \omega}{\beta \hbar \omega}
\end{align}
It follows from the exact expression for the density matrix of the harmonic oscillator (see \cite{Feynman:100771} and \cite{barragan2018harmonic}) that this effective potential gives the exact quantum partition function for the harmonic oscillator. \\

Similarly, the refined effective potential \eqref{VLH_final} obtained after renormalising \eqref{Veff_prelim} using the scaling factor \eqref{ZNCF} and applying the harmonic mapping substitution \eqref{magic} also yields the right classical limit
\begin{align}
    V_{\text{LH}}(x_a) \rightarrow V(x_a)
\end{align}
For a harmonic oscillator, the expression \eqref{PLH_final} is only equal to the exact quantum PDF upto a constant factor. However, this can be tackled by a numerical normalisation in simulations. 

\section{Evaluating Feynman-Hibbs and Feynman-Kleinert effective potentials}\label{appendix:FH and FK evaluation}
For a polynomial potential $V(x) = x^n$, the smeared potential 
\begin{align}
    V_{a^2} (\bar{x})  = \frac{1}{\sqrt{2\pi a^2}} \int_{-\infty}^{\infty} dx \: e^{-\frac{(x-\bar{x})}{2a^2}} V(x)
\end{align}
can be evaluated as:
\begin{align}
    V_{a^2} (\bar{x})  &= \frac{1}{\sqrt{2\pi a^2}} \int_{-\infty}^{\infty} dx \: e^{-x^2/2a^2} (x+\bar{x})^n \\
    &= \frac{1}{\sqrt{2\pi a^2}} \sum_{k=0}^{n} 
    \begin{pmatrix}
        n \\ k
    \end{pmatrix} \bar{x}^{n-k}
    \int_{-\infty}^{\infty} dx \: e^{-x^2/2a^2} x^k\\
    &=  \frac{1}{\sqrt{2\pi a^2}} \sum_{p=0}^{\floor{n/2}} 
    \begin{pmatrix}
        n \\ 2p
    \end{pmatrix} \bar{x}^{n-2p}
    \int_{-\infty}^{\infty} dx \: e^{-x^2/2a^2} x^{2p} \\
    &=\sum_{p=0}^{\floor{n/2}} 
    \begin{pmatrix}
        n \\ 2p
    \end{pmatrix}  (2p-1)!! a^{2p} \bar{x}^{n-2p}
\end{align}
Using this, we can directly evaluate the Feynman-Hibbs potential by setting $a^2 = \beta\hbar^2/3m$.
For the Feynman-Kleinert effective potential \eqref{eq:WFK}, we also need the variational parameter $\Omega$, which can be evaluated as:
\begin{equation}
\begin{split}
\Omega^2(x_0) = \sum_{p=0}^{\floor{n/2}-1} &\begin{pmatrix}
        n \\ 2p
    \end{pmatrix}  (2p-1)!! a^{2p} \\ &(n-2p) (n-2p-1)\bar{x}^{n-2p-2} 
\end{split}
\end{equation}

\bibliographystyle{aipnum4-2}
\bibliography{Citations}
\end{document}


\maketitle

\vspace{-1cm}

\begin{figure}
    \centering
    \includegraphics[scale=0.99]{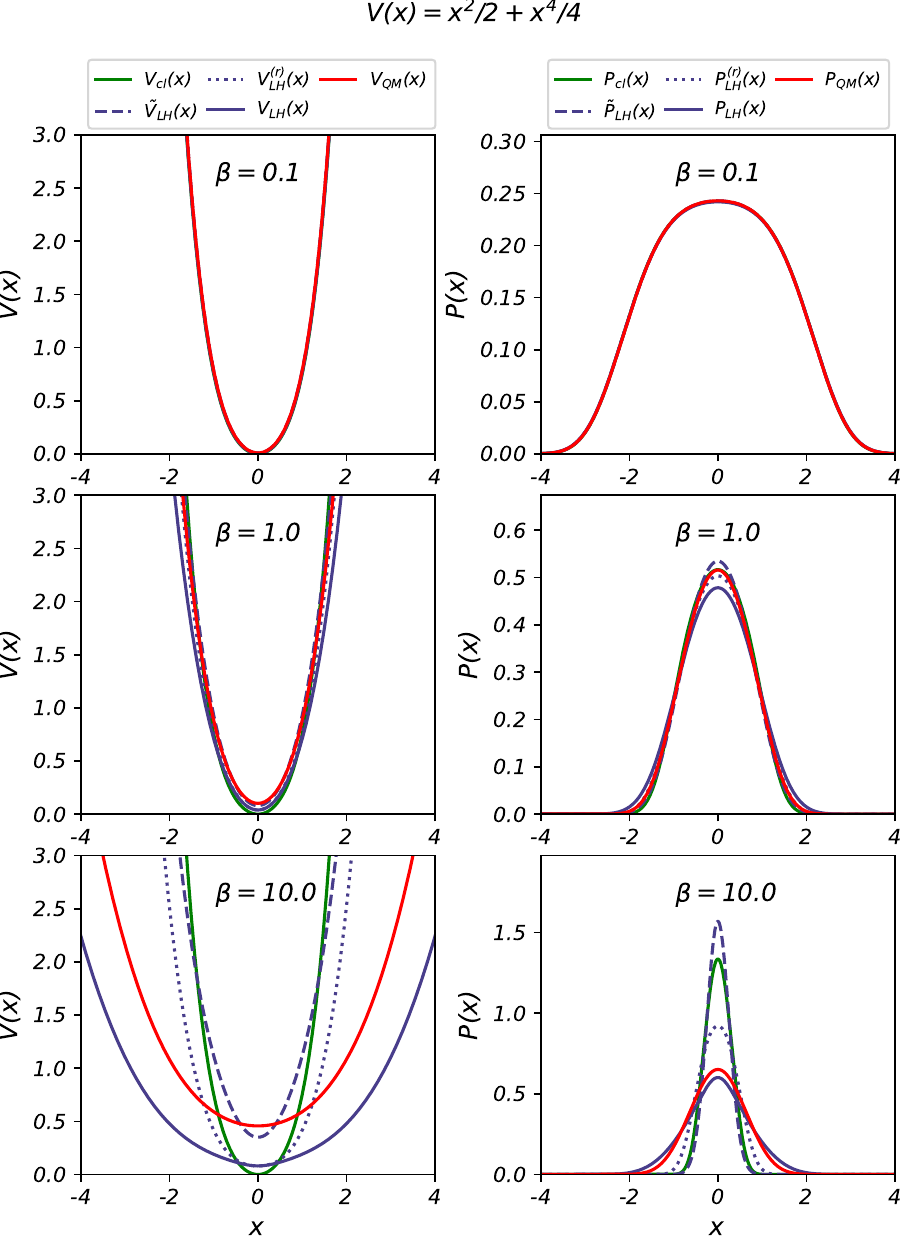}
    \caption{Left: Plots of the classical (i.e. $V(x)$), quantum and different forms of local harmonic effective potential for the quartic oscillator $V(x) = x^2/2 + x^4/4$ with $m=1$ at different temperatures. The functional form for each of these potentials are given in Table \ref{tab:1}. The finite shift in the energy at $\beta=10.0$ is due to the renormalisation with the non-classical factor \eqref{ZNCF}. Right: Plots of the corresponding position distribution functions as in Table \ref{tab:2}. The shift from \eqref{ZNCF} does not affect the PDF.  }
    \label{fig:harmquart}
\end{figure}

\begin{figure}
    \centering
    \includegraphics[scale=0.99]{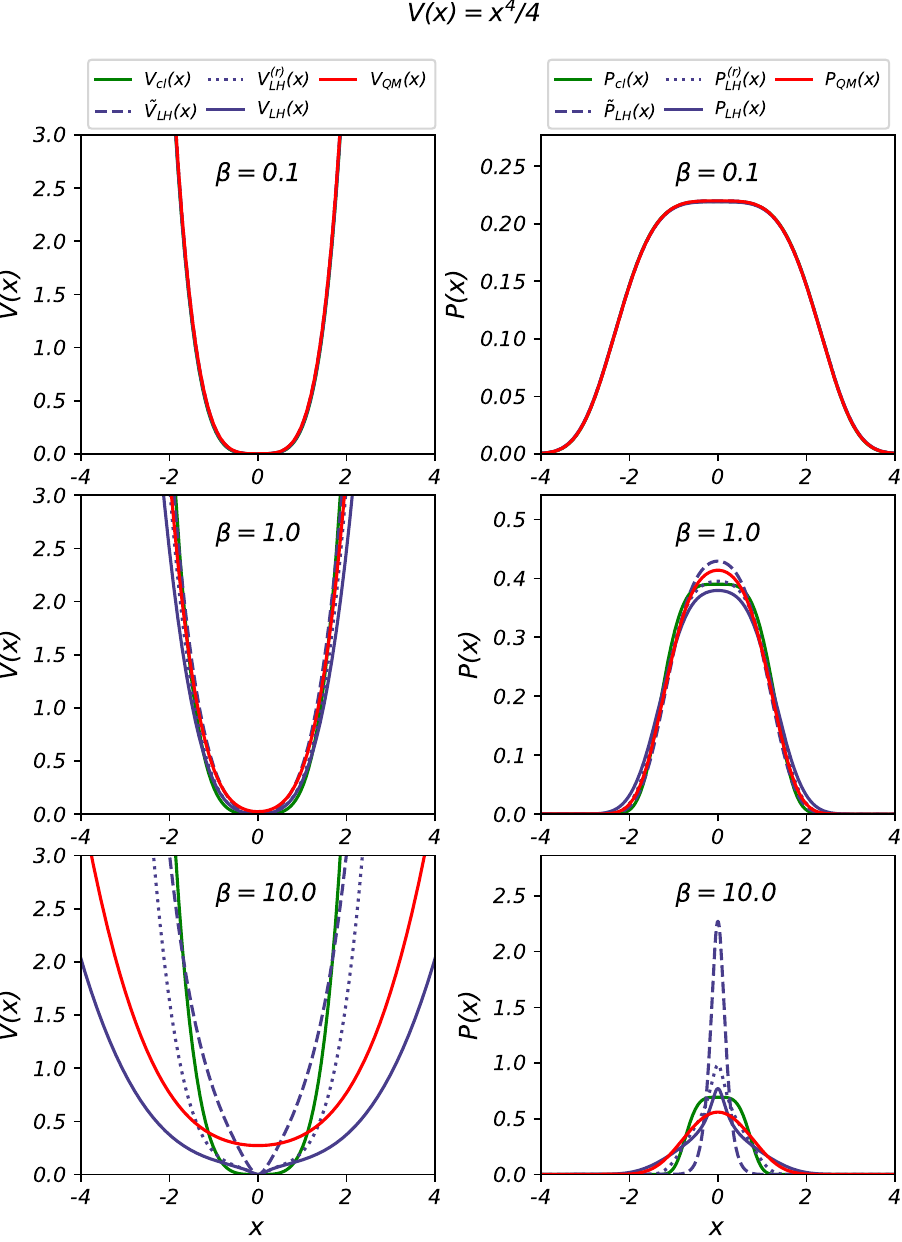}
    \caption{Same as in fig.~\ref{fig:harmquart}, but for a quartic oscillator $V(x) = x^4/4$. At low temperatures ($\beta=10.0$), the `bare' local harmonic effective potential \eqref{Vbare} is dominated by the term \eqref{Vnc2} which is linear in $\omega_a$ as $\beta\rightarrow \infty$ and causes a strong localisation around $x_a = 0$ where the curvature is $0$. The harmonic mapping substitution \eqref{magic} partially remedies this unphysical localisation at least for $x_a>0$. }
    \label{fig:quart}
\end{figure}

\begin{figure}
    \centering
    \includegraphics[scale=0.99]{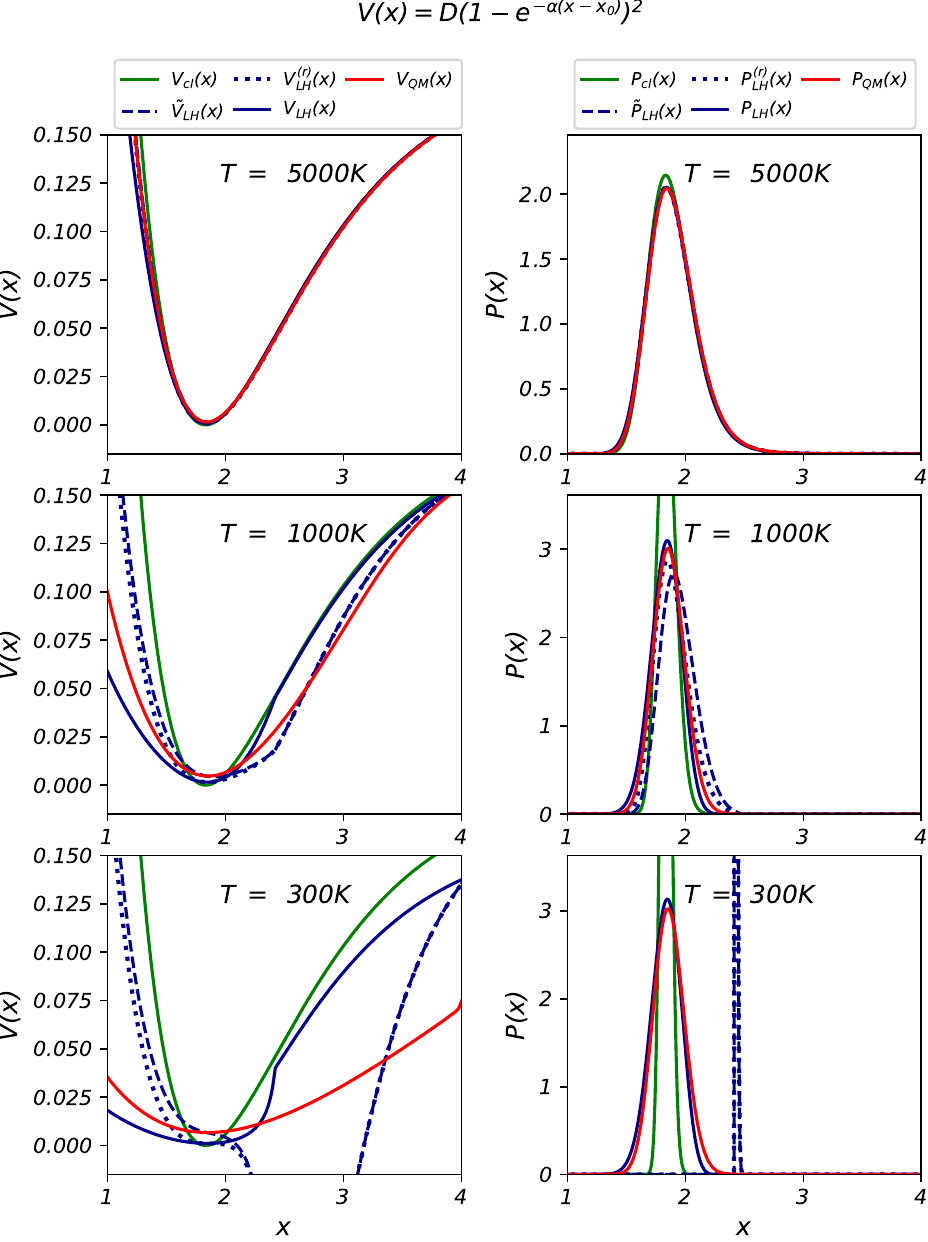}
    \caption{Same as in fig.~\ref{fig:morse}, but for a Morse oscillator $V(x)=D (1-e^{-\alpha (x-x_e)})^2$ with parameters as in the main text. For this system, the main contribution to \eqref{Vbare} at low temperatures ($T=300K$) is from \eqref{Vnc1} around the region $x \approx 2.5$ where the curvature of the Morse oscillator approaches $0$. This creates an artificial `well' around that point which causes unphysical localisation of the unnormalised distribution function. The refined potential \eqref{VLH_final} completely avoids this artificial well thanks to the substitution \eqref{magic}. }
    \label{fig:morse}
\end{figure}

We show the improvement in the position distribution function following the renormalisation of the `bare' local-harmonic effective classical potential by the non-classical factor and subsequent modification by the harmonic mapping substitution. Starting with the expression
\begin{align}\label{Vbare}
    \tilde{V}_{\text{LH}}(x_a) = V(x_a)  +
     \frac{m k_a^2}{2\omega_a^2}  \left(\frac{\tanh \xi_a}{\xi_a} -1\right) + \frac{1}{2\beta}\ln \frac{\sinh 2\xi_a}{2\xi_a}
\end{align}
we set
\begin{align}\label{Vnc1}
    \tilde{V}_{\text{LH}, \:nc1}(x_a) = 
     \frac{m k_a^2}{2\omega_a^2}  \left(\frac{\tanh \xi_a}{\xi_a} -1\right)
\end{align}
and 
\begin{align}\label{Vnc2}
        \tilde{V}_{\text{LH}, \:nc2}(x_a) = \frac{1}{2\beta}\ln \frac{\sinh 2\xi_a}{2\xi_a}
\end{align}
The refined expression (following renormalisation and harmonic mapping substitution) for the local-harmonic effective potential as in the main text is:
\begin{align}\label{VLH_final}
    V_{\text{LH}}(x_a) = V(x_a)\frac{\tanh \xi_a}{\xi_a}- \frac{1}{2\beta} \ln \frac{\tanh \xi_a}{\xi_a}
\end{align}
We also consider the position distribution function $P$ for the obtained from \eqref{Vbare} and \eqref{VLH_final} following a numerical normalisation. For reference, we plot the classical, quantum and `renormalised' distribution  
\begin{equation}\label{Prenorm}
\begin{split}
    {P}^{(r)}_{\text{LH}}(x_a) 
    =&\sqrt{\frac{m}{2 \pi\beta\hbar^2}} \sqrt{\frac{\tanh \xi_a}{\xi_a}}  \exp \left \{-\beta \left (V(x_a) - \frac{m k_a^2}{2\omega_a^2}  \right) \right \} \\&\times\exp\left\{- \beta\frac{m k_a^2}{2\omega_a^2} \left( \frac{\tanh \xi_a}{\xi_a} \right)\right\} 
\end{split}
\end{equation}
before applying the harmonic mapping substitution 
\begin{align}\label{magic}
    mk_a^2/(2\omega_a^2) \rightarrow V(x_a),
\end{align}
obtained by scaling the `bare' distribution function (from \eqref{Vbare}) by the non-classical factor
\begin{align}\label{ZNCF}
      {\Upsilon_{\text{NCF}}(x_a)} = \frac{Z_{\text{QM,local}}(x_a)}{Z_{\text{cl,local}}(x_a)} = \frac{\xi_a}{\sinh \xi_a}.
\end{align}
The corresponding effective potential is:
\begin{align}
    V_{\text{LH}}^{(r)}(x_a) = V(x_a) + \frac{m k_a^2}{2\omega_a^2}  \left(\frac{\tanh \xi_a}{\xi_a} -1\right) - \frac{1}{2\beta} \ln \frac{\tanh \xi_a}{\xi_a} 
\end{align}
The functional forms for the different effective potentials and the respective position distribution functions are summarised in Table \ref{tab:1} and \ref{tab:2} respectively. The results for perturbed harmonic, quartic and morse oscillator are shown in fig.~\ref{fig:harmquart}, fig.~\ref{fig:quart}
and fig.~\ref{fig:morse} respectively.

\renewcommand{\arraystretch}{2} 
\begin{table*}[t]
  \centering
  \begin{tabular}{lcr}
    Potential & Functional form  \\
    $V_{\text{cl}}(x_a) $ & $V(x_a)$\\
    $\tilde{V}_{\text{LH}}(x_a)$ &  $ V(x_a)  +
     \frac{m k_a^2}{2\omega_a^2}  \left(\frac{\tanh \xi_a}{\xi_a} -1\right) + \frac{1}{2\beta}\ln \frac{\sinh 2\xi_a}{2\xi_a}$ \\
     $V_{\text{LH}}^{(r)}(x_a)$ & $V(x_a) + \frac{m k_a^2}{2\omega_a^2}  \left(\frac{\tanh \xi_a}{\xi_a} -1\right) - \frac{1}{2\beta} \ln \frac{\tanh \xi_a}{\xi_a}$\\
     $ V_{\text{LH}}(x_a)$ & $ V(x_a)\frac{\tanh \xi_a}{\xi_a}- \frac{1}{2\beta} \ln \frac{\tanh \xi_a}{\xi_a}$\\
     $ V_{\text{QM}}(x_a)$ & $ -\beta^{-1} \log \bra{x_a} e^{-\beta\hat{H}}\ket{x_a} + \frac{1}{2\beta} \ln \frac{m}{2\pi\beta\hbar^2}$
  \end{tabular}
  \caption{Classical, quantum and local harmonic effective potentials}
  \label{tab:1}
\end{table*}

\renewcommand{\arraystretch}{2} 
\begin{table*}[t]
  \centering
  \begin{tabular}{ccc}
    Distribution Function & Functional form  \\
    $P_{\text{cl}}(x_a) $ & $\frac{1}{Z_{\text{cl}}}
    \sqrt{\frac{m}{2\pi\beta\hbar^2}}e^{-\beta V(x_a)}$\\
    $\tilde{P}_{\text{LH}}(x_a)$ &  $ \sqrt{\frac{m}{2\pi\beta\hbar^2}}\sqrt{\frac{2\xi_a}{\sinh 2\xi_a}}   \exp \left \{-\beta \left (V(x_a) + \frac{m k_a^2}{2\omega_a^2} \left( \frac{\tanh \xi_a}{\xi_a} - 1 \right)\right) \right \} $ \\
     $P_{\text{LH}}^{(r)}(x_a)$ & $\sqrt{\frac{m}{2 \pi\beta\hbar^2}} \sqrt{\frac{\tanh \xi_a}{\xi_a}}  \exp \left \{-\beta \left (V(x_a) - \frac{m k_a^2}{2\omega_a^2}  \right)  - \beta\frac{m k_a^2}{2\omega_a^2} \left( \frac{\tanh \xi_a}{\xi_a} \right) \right\}$  \\
     $ P_{\text{LH}}(x_a)$ & $\sqrt{\frac{m}{2 \pi\beta\hbar^2}} \sqrt{\frac{\tanh \xi_a}{\xi_a}} \exp \left \{-\beta V(x_a) \frac{\tanh \xi_a}{\xi_a}\right\}$ \\
     $ P_{\text{QM}}(x_a)$ & $ \frac{1}{Z_{\text{QM}}}  \bra{x_a} e^{-\beta\hat{H}}\ket{x_a} $
  \end{tabular}
  \caption{Classical, quantum and local harmonic position distribution functions (up to a constant numerical normalisation factor).}
  \label{tab:2}
\end{table*}

\newpage

